\documentclass[aps,prl,twocolumn,superscriptaddress,reprint,showpacs]{revtex4-1}
\usepackage{graphicx}
\usepackage{setspace}
\usepackage{amsmath}
\usepackage{amssymb}
\usepackage{color}
\usepackage{array}
\usepackage{subfigure}
\usepackage{float}
\usepackage{lipsum}
\usepackage[
   bookmarks=true,
   colorlinks=true,
   hyperindex=true,
   citecolor=blue,
   linkcolor=blue,
   urlcolor=blue
]{hyperref}
\usepackage{etoolbox}
\usepackage{placeins}
\patchcmd{\thebibliography}{\section*{\refname}}{}{}{}

\begin{document}
\title{A fiber array architecture for atom quantum computing}

\author{Xiao Li}
\affiliation{Division of Precision Measurement Physics, Innovation Academy for Precision Measurement Science and Technology, Chinese Academy of Sciences, Wuhan 430071, China}

\author{Jia-Yi Hou}
\affiliation{Division of Precision Measurement Physics, Innovation Academy for Precision Measurement Science and Technology, Chinese Academy of Sciences, Wuhan 430071, China}
\affiliation{School of Physical Sciences, University of Chinese Academy of Sciences, Beijing 100049, China}

\author{Jia-Chao Wang}
\affiliation{Division of Precision Measurement Physics, Innovation Academy for Precision Measurement Science and Technology, Chinese Academy of Sciences, Wuhan 430071, China}
\affiliation{School of Physical Sciences, University of Chinese Academy of Sciences, Beijing 100049, China}

\author{Guang-Wei Wang}
\affiliation{Division of Precision Measurement Physics, Innovation Academy for Precision Measurement Science and Technology, Chinese Academy of Sciences, Wuhan 430071, China}
\affiliation{School of Physical Sciences, University of Chinese Academy of Sciences, Beijing 100049, China}

\author{Xiao-Dong He}
\affiliation{Division of Precision Measurement Physics, Innovation Academy for Precision Measurement Science and Technology, Chinese Academy of Sciences, Wuhan 430071, China}
\affiliation{Wuhan Institute of Quantum Technology, Wuhan 430206, China}

\author{Feng Zhou}
\affiliation{Division of Precision Measurement Physics, Innovation Academy for Precision Measurement Science and Technology, Chinese Academy of Sciences, Wuhan 430071, China}

\author{Yi-Bo Wang}
\affiliation{Division of Precision Measurement Physics, Innovation Academy for Precision Measurement Science and Technology, Chinese Academy of Sciences, Wuhan 430071, China}

\author{Min Liu}
\affiliation{Division of Precision Measurement Physics, Innovation Academy for Precision Measurement Science and Technology, Chinese Academy of Sciences, Wuhan 430071, China}

\author{Jin Wang}
\affiliation{Division of Precision Measurement Physics, Innovation Academy for Precision Measurement Science and Technology, Chinese Academy of Sciences, Wuhan 430071, China}
\affiliation{Wuhan Institute of Quantum Technology, Wuhan 430206, China}
\affiliation{Hefei National Laboratory, Hefei 230088, China}

\author{Peng Xu}
\email{etherxp@wipm.ac.cn}
\affiliation{Division of Precision Measurement Physics, Innovation Academy for Precision Measurement Science and Technology, Chinese Academy of Sciences, Wuhan 430071, China}
\affiliation{Wuhan Institute of Quantum Technology, Wuhan 430206, China}

\author{Ming-Sheng Zhan}
\email{mszhan@apm.ac.cn}
\affiliation{Division of Precision Measurement Physics, Innovation Academy for Precision Measurement Science and Technology, Chinese Academy of Sciences, Wuhan 430071, China}
\affiliation{Wuhan Institute of Quantum Technology, Wuhan 430206, China}
\affiliation{Hefei National Laboratory, Hefei 230088, China}

\begin{abstract}
Arrays of single atoms trapped in optical tweezers are increasingly recognized as a promising platform for scalable quantum computing. In both the fault-tolerant and NISQ eras, the ability to individually control qubits is essential for the efficient execution of quantum circuits. Time-division multiplexed control schemes based on atom shuttling or beam scanning have been employed to build programmable neutral atom quantum processors, but achieving high-rate, highly parallel gate operations remains a challenge. Here, we propose a fiber array architecture for atom quantum computing capable of fully independent control of individual atoms. The trapping and addressing lasers for each individual atom are emitted from the same optical waveguide, enabling robust control through common-mode suppression of beam pointing noise. Using a fiber array, we experimentally demonstrate the trapping and independent control of ten single atoms in two-dimensional optical tweezers, achieving individually addressed single-qubit gate with an average fidelity of 0.9966(3). Moreover, we perform simultaneous arbitrary single-qubit gate on four randomly selected qubits, resulting in an average fidelity of 0.9961(4). Our work paves the way for time-efficient execution of quantum algorithms on neutral atom quantum computers.

\end{abstract}

\maketitle

\section{Introduction}
Owing to the outstanding scalability in qubit numbers \cite{Antoine-PRX-SLM,endres2016atom,barredo2016atom,barredo2018synthetic,schlosser2023scalable,manetsch2024tweezer} and reconfigurable Rydberg interactions \cite{ebadi2021quantum,scholl2021quantum,ebadi2022quantum,eckner2023realizing,bornet2023scalable,cao2024multi,finkelstein2024universal}, single-atom arrays trapped in optical tweezers serves as a crucial platform for exploring new physics in complex quantum systems \cite{browaeys2020many,altman2021quantum,bernien2017probing,de2019observation,kim2020quantum}, while also being a prominent candidate for quantum computing \cite{bluvstein2022quantum,graham2022multi, bluvstein2024logical, radnaev2024universal, young2024atomic,yang2016coherence,tian2023coherence,barnes2022assembly, xia2015randomized, wang2016single, sheng2018high, nikolov2023randomized, levine2019parallel, fu2022high, evered2023high, ma2023high, scholl2023erasure}. Recent experiments have demonstrated logical qubit encoding and logical gate operations on hundreds of moving atoms \cite{bluvstein2024logical}, emphasizing the notable advantages of the atom array platform in terms of qubit count, gate operation parallelism, and qubit connectivity. However, achieving practical quantum computing still requires significant progress in reducing error rates and increasing clock speeds \cite{beverland2022assessing,poole2024architecture, pecorari2024high}. Thus, one of the upcoming key challenges for atom quantum computing is to develop an architecture that supports fast, highly parallel, scalable and stable addressing operations.

A natural way to achieve programmable quantum computing on this platform is to first use optical diffraction devices, such as spatial light modulators (SLMs) \cite{barredo2016atom} or acousto-optic deflectors (AODs) \cite{endres2016atom}, to create large qubit arrays, and then incorporate individual addressing capabilities into these arrays. In recent years, two schemes based on time-division multiplexed optical addressing have been demonstrated to realize this architecture. One involves multiplexing the addressing light by shuttling atoms, where qubit addressing is achieved by shifting qubits in and out of a large, uniform driving laser \cite{bluvstein2022quantum,bluvstein2024logical}. This facilitates parallel qubit operations and enables on-demand, non-local qubit connections. However, moving qubits spatially, especially between different trap sites, will significantly increase the idle time between qubit gate operations, typically reaching several hundred microseconds. Another approach uses an AOD to steer tightly focused addressing beams onto a static single-atom array, reducing the idle time to sub-microseconds \cite{graham2022multi,radnaev2024universal}. However, this scheme favors sequential operations and faces challenges in maintaining precise and stable alignment of the addressing light with the qubits. Additionally, both approaches use AODs to achieve fast atom movement or quick beam steering, which restricts the simultaneously addressed qubits to a square grid pattern. In the long term, executing deep quantum circuits on a large number of qubits is necessary for practical error-correcting quantum algorithms. This requires high gate rates and highly parallel operations, considering finite coherence times and constraints on acceptable execution times \cite{beverland2022assessing}. In the NISQ era, high flexibility in qubit implementation is crucial for achieving quantum advantage, such as in random quantum circuit sampling \cite{arute2019quantum,wu2021strong}. Several other addressing schemes have been proposed, but fully meeting the numerous requirements remains challenging \cite{menssen2023scalable,zhang2024scaled,graham2023multiscale}.

\begin{figure*}
   \includegraphics[width=0.8\linewidth]{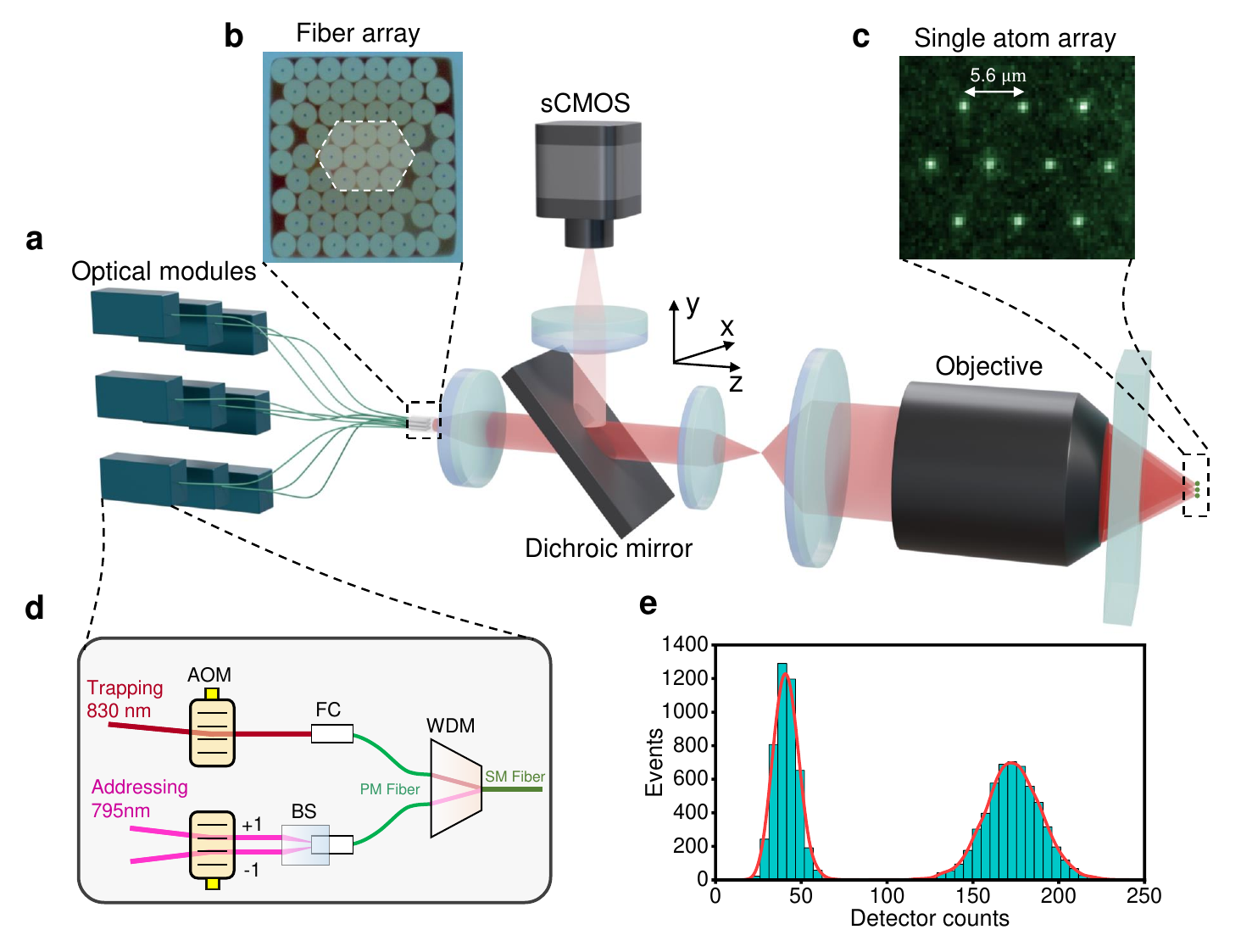}
   \caption{\textbf{Experimental scheme.} \textbf{a}, Basic experimental setup for the trapping, manipulation, and detection of single atom arrays. Each optical module is used for trapping and addressing an individual atom. The optical tweezers are generated using an optical fiber array. Multi-channel trapping beams are emitted from the end face of the fiber array and, after collimation and expansion, are focused by a objective (NA = 0.7) into the vacuum chamber, creating an optical trap array. The addressing beams are focused onto the trap array along the same optical path as the trapping beams. Atom fluorescence signals are collected by the same objective and imaged onto an sCMOS camera after being reflected by a dichroic mirror. \textbf{b}, The cross-section of the fiber array consists of 64 single-mode fibers, each with cladding reduced to $40\,\mu\text{m}$ at the ends. The 10 fibers enclosed in the dashed box are chosen for this experiment. \textbf{c}, Averaged fluorescence images of the single-atom array, taken with an exposure time of 50 ms. \textbf{d}, A schematic showing optical setup in the optical module. A single $830\,$nm trapping beam and a pair of $795\,$nm addressing Raman beams each pass through an AOM before being combined into the same SM fiber using a WDM. \textbf{e}, Histogram of collected photons for one of the traps. Two distinct peaks indicate the presence of one atom (right peak) and no atom (left peak) in the trap.
   \label{fig1}}
\end{figure*}

To overcome these challenges, we propose a modular fiber array architecture to independently control single-atom qubits in atom arrays for quantum computing. In each module, the trapping and addressing laser beams are combined into a single channel of an fiber array. Then, the beams are focused on the same location through identical optical paths, enabling stable and full control of individual atoms. The number of atoms and control signals can be directly scaled by replicating the optical modules and waveguide channels. Experimentally, we used a fiber array to trap and independently control 10 single atoms, demonstrating individually addressed single-qubit gates with an average fidelity of 0.9966(3). Notably, we also achieved parallel implementation of arbitrary single-qubit gates on this 2D atom array, a task previously considered highly challenging. Our work presents a robust method for the independently control of qubits in a two-dimensional single-atom array. Combined with high fidelity addressed long-range Rydberg gates \cite{radnaev2024universal,poole2024architecture, pecorari2024high}, this offers a feasible architecture for executing quantum algorithms on static single-atom qubits with high time efficiency.

\section{Results}

\subsection{Scheme}

In our previous design \cite{ke2016tailored}, lasers emitted from a one dimensional waveguide array were directly focused by a microlens array, thereby creating an array of light spots. In order to allow adjacent single atoms to enter the Rydberg blockade region, the spacing between the optical waveguides needs to be reduced to just a few micrometers, leading to substantial crosstalk between waveguides. Moreover, etching microlenses at the ends of optical waveguides introduces considerable aberrations.

In this work, we utilize a 2D fiber array to provide transmission channels for both trapping light and addressing light. The basic architecture is shown in Fig.\,\ref{fig1}\textbf{a}. The light field, emitted from a fiber array with spacing of several tens of micrometers, are projected inside a vacuum chamber with a reduced image, forming an array of optical traps spaced just a few micrometers apart. This configuration relaxes the spacing requirements between optical waveguide channels to tens of micrometers. Moreover, optical aberrations during beam focusing can be effectively corrected with the advanced optical design techniques. It should be noted that similar methods based on 1D waveguide arrays have also been recently employed to address single ions in linear ion traps \cite{pogorelov2021compact,binai2023guided,sotirova2024low,flannery2024physical}.

The experimental setup is shown in Fig.\,\ref{fig1}\textbf{a}. We customize a 2D fiber array consisting of 64 single-mode fibers (Nufern 780-HP fiber), with the cladding at the end of each single-mode fiber reduced to $40\,\mu$m. These fibers are then inserted and bonded into a glass capillary tube. The cross-section of the fiber array, as illustrated in Fig.\,\ref{fig1}\textbf{b}, shows that most of the single-mode fibers arranged tightly. We select 10 tightly packed single-mode fibers (as indicated by the dashed box in Fig.\,\ref{fig1}\textbf{b}) to create 10 optical dipole traps.

In addition to atom trapping, this fiber array also enables individual addressing within the atom arrays. The trapping and addressing light for each single atom are contained within a single optical module, as shown in Fig.\,\ref{fig1}\textbf{d}. In each module, the $830\,$nm trapping beam passes through an acousto-optic modulator (AOM), with the first-order diffracted laser coupled into a single-mode polarization-maintaining fiber. Two $795\,$nm Raman addressing beams pass through a single AOM, producing +1 and -1 order diffracted beams that are subsequently combined into a single-mode polarization-maintaining fiber (see the next section for details). The diffracted trapping and addressing beams are finally combined into one of the single-mode fibers of the fiber array through a wavelength-division multiplexer (WDM). To improve the stability of the optical setup and reduce its volume, we employ adhesive bonding techniques to construct 10 modules. Additionally, the fiber array is mounted on a fixed rail to ensure the polarization stability of the laser transmitted within it.

The 10 trapping laser beams are individually coupled into 10 selected single-mode fibers arranged in a hexagonal pattern and emit from the end face of the fiber array, forming 10 Gaussian beams with a spacing of $40\,\mu\text{m}$. After being collimated and expanded, these trapping beams are focused into the vacuum chamber by a microscope objective with a numerical aperture (NA) of 0.7, resulting in an optical trap array with a beam waist of about $0.65\,\mu$m. The magnification of the optical system is 0.14, so the spacing of the optical trap array is around $5.6\,\mu$m. Ten individual Rubidium-87 atoms are successfully loaded from a 3D-MOT into these optical tweezers, with a loading rate of 0.55. Fig.\,\ref{fig1}\textbf{c} shows the averaged atomic fluorescence image of the tweezer array, with an exposure time of 50 ms. Fig.\,\ref{fig1}\textbf{e} shows the histogram of collected photons for one of the traps, where two distinct peaks indicating the presence of one atom (right peak) and no atom (left peak) in the trap.

\subsection{Individual addressing of single-atom qubit arrays}
\begin{figure}
   \includegraphics[width=\linewidth]{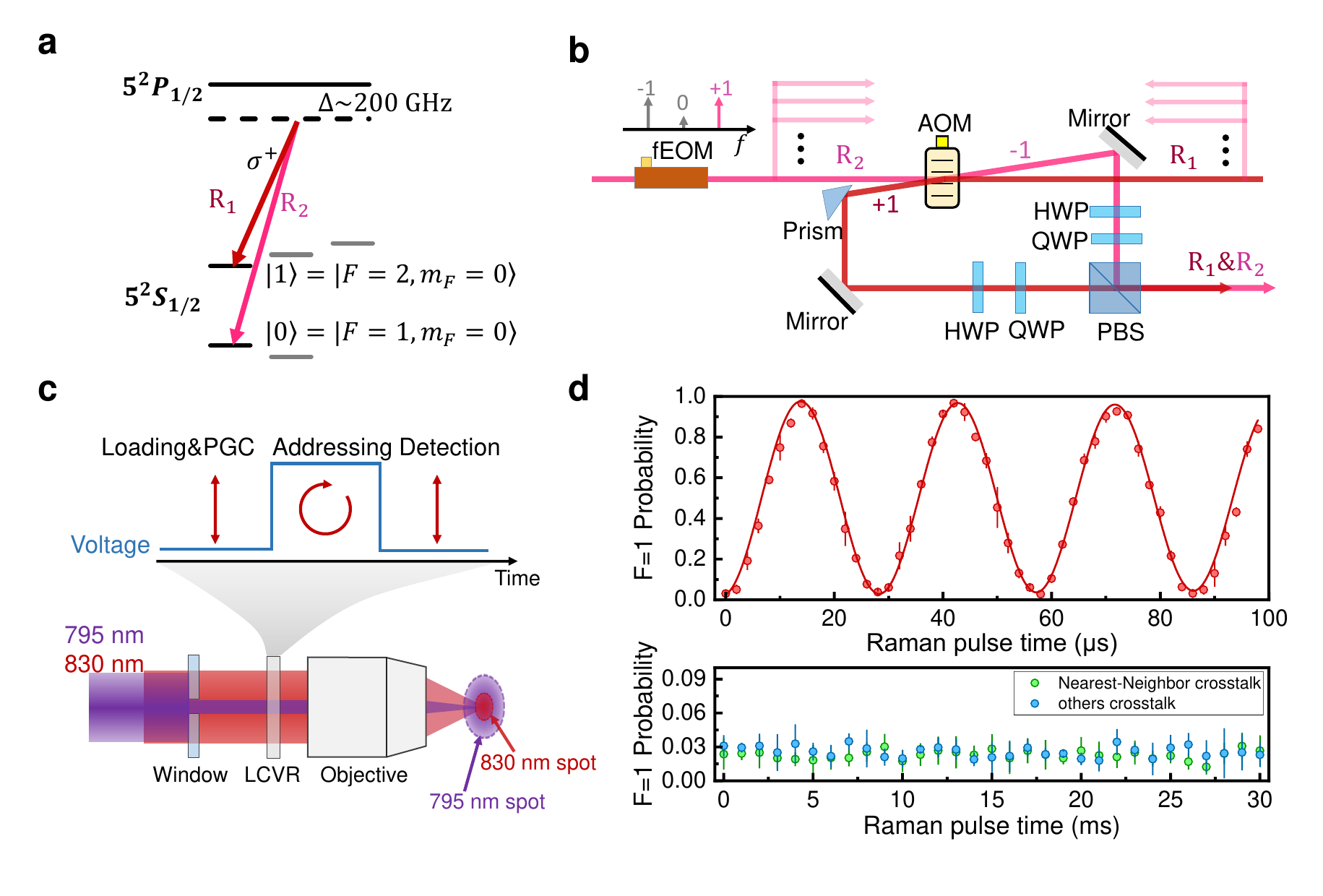}
   \caption{\textbf{Individual addressing of single-atom qubit arrays.} \textbf{a}, Energy level schematic for the qubits. Qubits are encoded in the two hyperfine ground states of ${}^{87}\mathrm{Rb}$ atoms: $|0\rangle\equiv |5S_{1/2},F=2, m_F=0\rangle$ and $|0\rangle\equiv |5S_{1/2},F=1, m_F=0\rangle$, which are coupled by a two-photon Raman transition featuring a single-photon detuning of $2\pi \times 200\,\text{GHz}$ relative to the $5P_{1/2}$ manifold. \textbf{b}, Optical configuration for Raman laser beams. Two coherent laser beams, one modulated by an fEOM and the other unmodulated, are each split into ten paths. Each pair of split beams ($R_1$ and $R_2$) pass through an AOM from opposite sides, obtaining +1 and -1 order diffractions respectively. They are then combined into a single SM PM fiber with identical polarization. The driving frequencies for the fEOM and AOM are set at 7.054 GHz and 100 MHz, respectively. The +1 sideband of the fEOM on $R_2$, in combination with $R_1$, forms a pair of Raman beams with a frequency difference of 6.834 GHz. \textbf{c}, Optical methods for switching laser polarization and expanding the addressing spot. The driving voltage amplitude of the LCVR is varied to meet the laser polarization requirements for different experimental phase. During the atom loading, PGC, and detection phases, the polarization is set to linear. In the addressing phase, it is switched to circular. A window featuring a 1 mm diameter central aperture, which is highly reflective at $795\,$nm and highly transmissive at $830\,$nm, is inserted into the optical path to enlarge the beam waist of the addressing spot. \textbf{d}, Rabi oscillation of the 10 single atoms when addressing one single atom. The upper panel displays the Rabi oscillations of a specific qubit (qubit 7), driven by a Raman pulse, whereas the lower panel depicts the evolution of other qubits subjected to a longer Raman pulse.
   \label{fig2}}
\end{figure}

We first demonstrate individual addressing of this atom array. Fig.\,\ref{fig2}\textbf{a} shows the relevant atomic-level diagram. Qubits are encoded in the two hyperfine ground states of ${}^{87}\mathrm{Rb}$ atoms: $|0\rangle\equiv |5S_{1/2},F=2, m_F=0\rangle$ and $|0\rangle\equiv |5S_{1/2},F=1, m_F=0\rangle$. Arbitrary rotations of each individual qubit are achieved via an addressed two-photon Raman transition. The Raman laser is set 200 GHz red-detuned to the transition between the $5S_{1/2}$ and $5P_{1/2}$ states.

In order to independently control 10 qubits, we need 10 sets of Raman laser modules with full control over amplitude, frequency, and phase. We develop a convenient and cost-effective solution, with the optical setup shown in Fig.\,\ref{fig2}\textbf{b}. The two Raman laser beams come from the same $795\,$nm cat-eye laser. One beam passes through a fiber-coupled electro-optic modulator (FEOM) for phase modulation, while the other does not. Firstly, the two Raman laser beams are split into ten separate paths. Each pair of split beams is then pass through a single AOM from opposite sides, producing +1 order (without FEOM) and -1 order (with FEOM) diffracted beams. After that, the two diffracted beams are combined into a single-mode polarization-maintaining fiber with the same polarization state, forming one pair of addressing Raman laser beams. For this experiment, the FEOM is modulated at 7.054 GHz, and the modulation frequency of the AOM is set around 110 MHz (each AOM's modulation frequency can be finely tuned according to the specific transition frequency of each single-atom qubit).

The 10 addressing beams are coupled into the 10 single-mode fibers, already carrying trapping light, through 10 separate WDMs. Each addressing beam emitted from the fiber array travels through the same optical setup as the trapping beam and is focused on each site in the single-atom array. Individual and parallel manipulation of arbitrary single-atom qubits can be achieved by controlling the amplitude, frequency, and phase of each addressing laser.

In our architecture, the trapping and addressing light share the same optical path after emerging from the fiber array. This enhances the stability and flexibility of addressing control but introduces challenges in differentiating the beam parameters for the trapping and addressing light.

The polarization of the laser light is the primary consideration. During atom loading and polarization gradient cooling (PGC), the trapping light needs to be linearly polarized to achieve low-temperature single atoms. For the site-selective control stage, circular polarization of the addressing light is required to efficiently drive the Raman transitions. We solve this contradiction by dynamically adjusting the laser polarization using a liquid crystal variable retarder (LCVR). By switching the driving voltage amplitude of the LCVR, the laser polarization requirements for different experimental stages are met, as shown in Fig.\,\ref{fig2}\textbf{c}. Moreover, this polarization setting also forms magic-intensity trapping for the atom array, thereby increasing the coherence time of the single-atom qubits \cite{yang2016coherence}.

Another aspect to take into account is that the alignment accuracy between the addressing laser beam spots and the single atoms, as well as the spatial uniformity of the addressing light experienced by the single atoms, will affect the addressing precision. For the first influencing factor, we choose achromatic singlet lenses and customize an apochromatic objective to ensure that the axial chromatic aberration of the optical system is less than $1\,\mu$m in the $795\,$nm (addressing laser wavelength) to $830\,$nm (trapping laser wavelength) range. For the second factor, there are two approaches: lowering the temperature of the single atoms and expanding the waist of the addressing beam spot, the latter being our choice in this work.

 To increase the addressing beam spot size, as shown in Fig.\,\ref{fig2}\textbf{c}, we insert a window with a 1 mm diameter central aperture before the dichroic mirror. The window is highly transmissive at $830\,$nm and highly reflective at $795\,$nm. The addressing laser beam, with an initial diameter of $4\,$mm, is reduced to $1\,$mm after passing through this window, while the trapping laser beam remains almost unaffected. Consequently, the waist of the addressing beam spot focused on the single atom increases from $0.65\,\mu$m to about $2\,\mu$m.

The experiment begins with single ${}^{87}\mathrm{Rb}$ atoms being randomly loaded into a two-dimensional optical tweezer array, with a loading rate of 0.55. The loading depth for each trap is 1.4 mK. The temperature of the single atoms is reduced to $18\,\mu$K after PGC. Next, with a $B \approx 3\,$G magnetic field (co-aligned with the direction of the dipole trap), the single atoms are pumped into $|0\rangle$ by $\pi$-polarized optical pumping light. Following this, we lower the trap depth to $200\,\mu$K and switch the polarizations of the dipole and addressing lights to $\sigma^+$ via an LCVR. In this circularly polarized optical dipole trap, the averaged coherence time of single-atom qubits can reach 50 ms. It should be noted that the frequencies of the trapping lasers at different sites are offset by more than 4 MHz to suppress atom heating and state leakage caused by interference between the trapping beams \cite{kobayashi2009fictitious}. Individual addressing on any single atom can be achieved by switching on the corresponding Raman laser beam.

The upper part of Fig.\,\ref{fig2}\textbf{d} exemplarily shows the Rabi oscillation between $|0\rangle$ and $|1\rangle$ on one single-atom qubit, driven by two-photon Raman transitions. The lower part displays the Rabi oscillations of qubits in other traps over an extended evolution time. It evident that the Rabi rate crosstalk (defined by \(\Omega_{\text{adjacent}}/\Omega_{\text{addressed}}\) \cite{pogorelov2021compact,li2023low}) on all other unaddressed single-atom qubits is less than $0.1\%$. Similarly, crosstalk for other target single atoms is measured, resulting in a maximum Rabi rate crosstalk of $1.0\%$ (see Methods for details).

\begin{figure}
   \includegraphics[width=\linewidth]{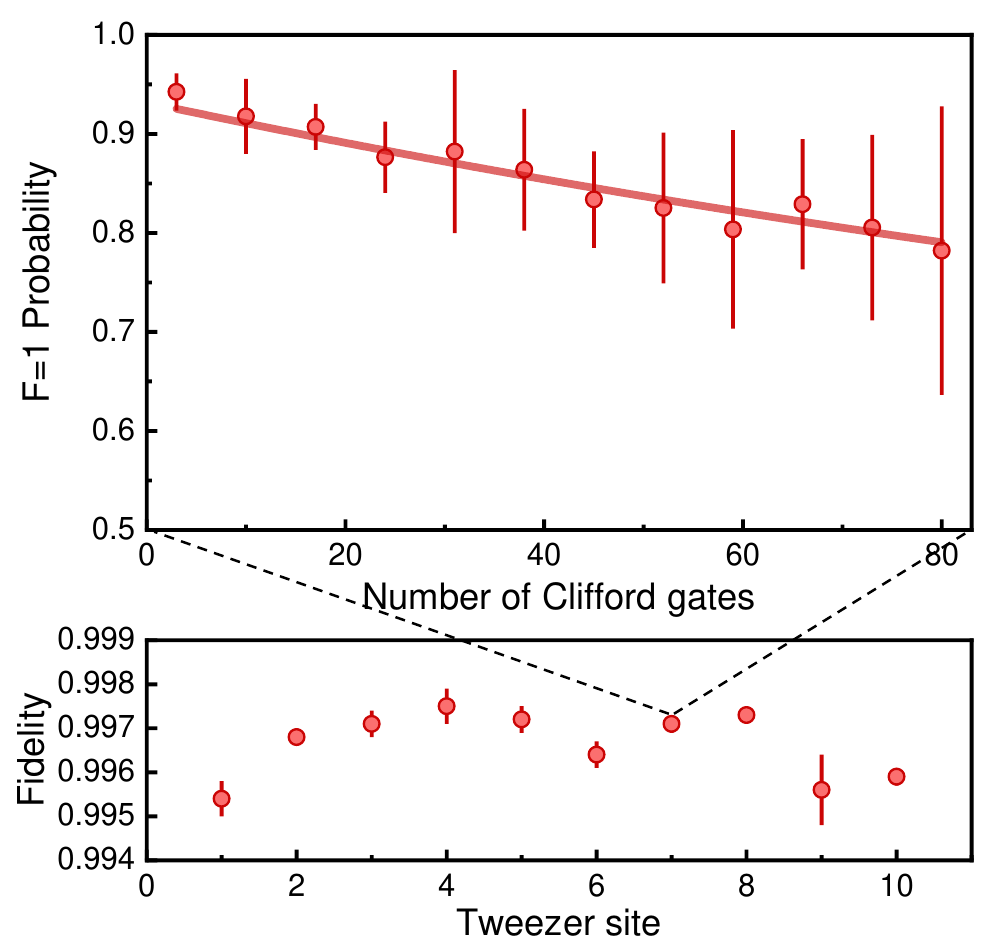}
   \caption{\textbf{RB of individually addressed single-qubit gates.} \text The upper panel shows RB experiment on one of the qubits (qubit 7), from which a fidelity of 0.9971 (2) for individually addressed single-qubit gates was extracted. The lower panel displays a site-by-site analysis of average single-qubit fidelities, which vary between 0.995 and 0.998.
   \label{fig3}}
\end{figure}

We evaluate the quality of the addressed single-qubit gates using the well-established Randomized Benchmarking (RB) method, based on random sampling from a set of 24 Clifford gates \cite{xia2015randomized,wang2016single,sheng2018high,nikolov2023randomized}. Each Clifford gate is generated from fundamental gates I, $R_{x,y,z}(\pm\pi/2)$, $R_{x,y,z}(\pi)$. Here, the Z gate is virtually implemented by switching the phase of the Raman laser \cite{nikolov2023randomized}. In the RB experiments, we generate 10 different random sequences of Clifford gates, truncated at various lengths. After each sequence, a correction gate is applied to rotate the qubit to $|1\rangle$. The fidelity of the addressed single-qubit gate can be extracted by fitting the equation below :
\begin{equation}
\bar{F} = \frac{1}{2} + \frac{1}{2}(1 - d_{if})(1 - 2\varepsilon_{g})^{\ell}, \label{eq1}
\end{equation}
where $\bar{F}$ is the average probability of $|1\rangle$ state, $\varepsilon_{g}$ represents the average error per Clifford gate, $d_{if}$ represents the depolarization probability associated with state preparation, $\ell$ denotes the number of random gates applied to the target qubit.

We perform RB experiments on the entire single-atom array site by site. Single-qubit RB data for one site are displayed in the upper panel of Fig.\,\ref{fig3}. The fidelity of addressed single-qubit gates at each site is summarized in the lower panel of Fig.\,\ref{fig3}, with the fidelity across the entire array ranging from 0.995 to 0.998 and an average fidelity of 0.9966(3). While a detailed error analysis is not performed, for our current experimental setup, gate errors mainly come from two sources: (1) Spatial variation in the Rabi frequency experienced by single atoms, and (2) the disruption of qubit coherence due to the spatially non-uniform differential AC Stark shift on hyperfine states of trapped single atoms, induced by addressing light. They can both be effectively suppressed by lowering the temperature of the single atoms.

\subsection{Parallel addressing of arbitrary single-atom qubits}
\begin{figure}
   \includegraphics[width=\linewidth]{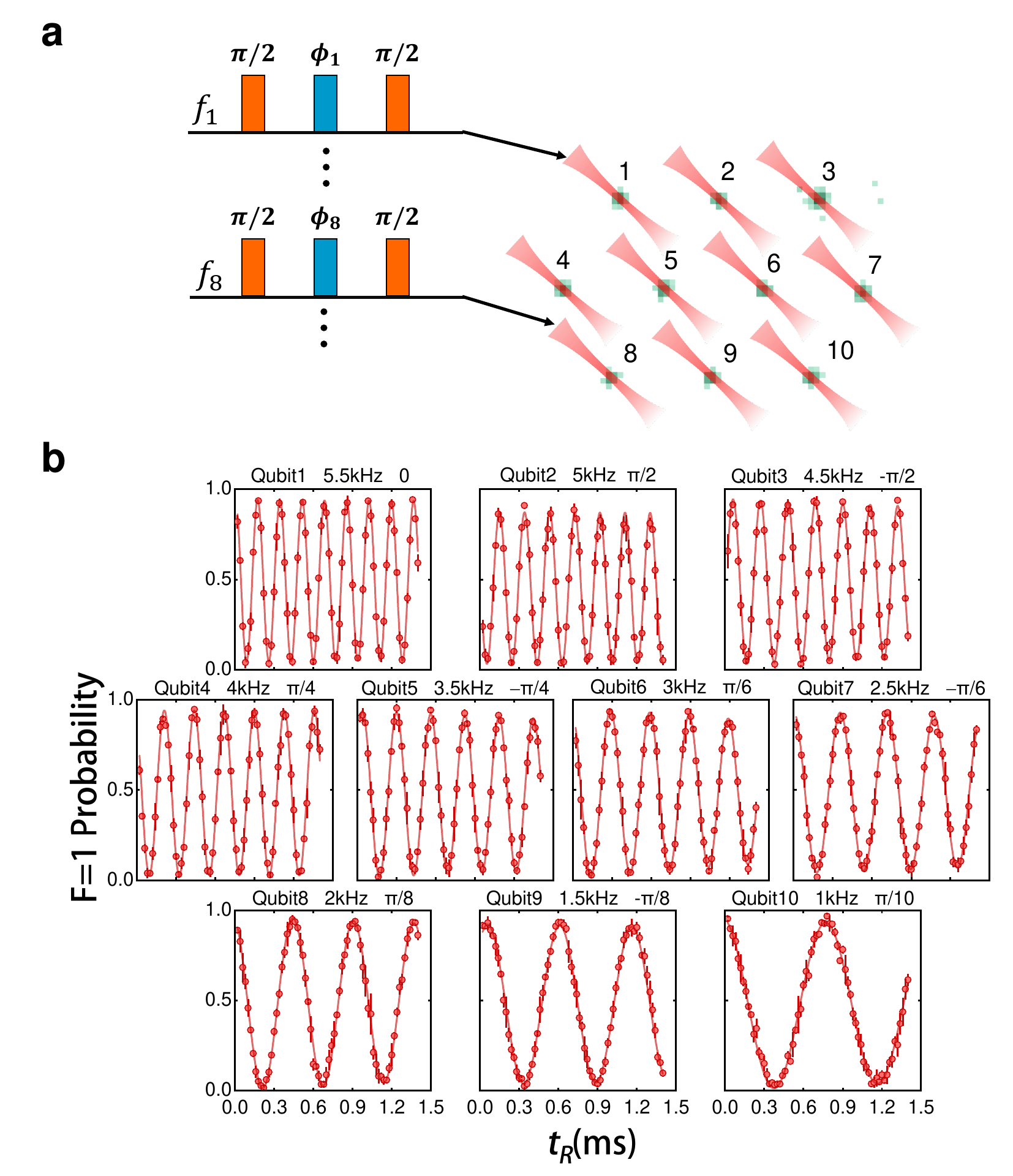}
   \caption{\textbf{Simultaneous Ramsey experiment on the entire array.} \textbf{a}, The Ramsey sequence comprises two $\pi/2$ pulses with a phase-shift pulse inserted between them to alter the phase of the second $\pi/2$ pulse. At each site, the Ramsey sequence is configured with unique two-photon frequencies $f_i$ and phase offsets $\phi_i$. \textbf{b}, 10 single-atom qubits undergo simultaneous Ramsey oscillations, each manifesting distinct frequencies \{5.5 kHz, 5 kHz, 4.5 kHz, 4 kHz, 3.5 kHz, 3 kHz, 2.5 kHz, 2 kHz, 1.5 kHz, 1 kHz\} and initial phases \{0, $\pi/2$, -$\pi/2$, $\pi/4$, $-\pi/4$, $\pi/6$, $-\pi/6$, $\pi/8$, $-\pi/8$, $\pi/10$\}.
   \label{fig4}}
\end{figure}

To demonstrate the ability in parallel manipulation of qubits, we first perform simultaneous Ramsey experiment on the entire atom array. The Ramsey sequence consists of three parts: two $\pi/2$ pulses separated by a waiting time, and an phase-shift pulse for varying the phase of the second $\pi/2$ pulse (see Methods). Here, we set unique detunings and phase offsets for the Ramsey sequence at each site, enabling each individual qubit to exhibit Ramsey oscillations with distinct frequencies and initial phases. From qubit 1 to qubit 10, detunings are evenly spaced from \(1 \text{ kHz}\) to \(5.5 \text{ kHz}\), and phase offsets are sequentially set to 0, $\pi/2$, -$\pi/2$, $\pi/4$, $-\pi/4$, $\pi/6$, $-\pi/6$, $\pi/8$, $-\pi/8$, $\pi/10$, as shown in Fig. \ref{fig4}\textbf{a}. Fig.\,\ref{fig4}\textbf{b} shows the results of applying Ramsey sequences simultaneously to 10 qubits, clearly illustrating the distinct Ramsey oscillations for each qubit.

\begin{figure}
   \includegraphics[width=\linewidth]{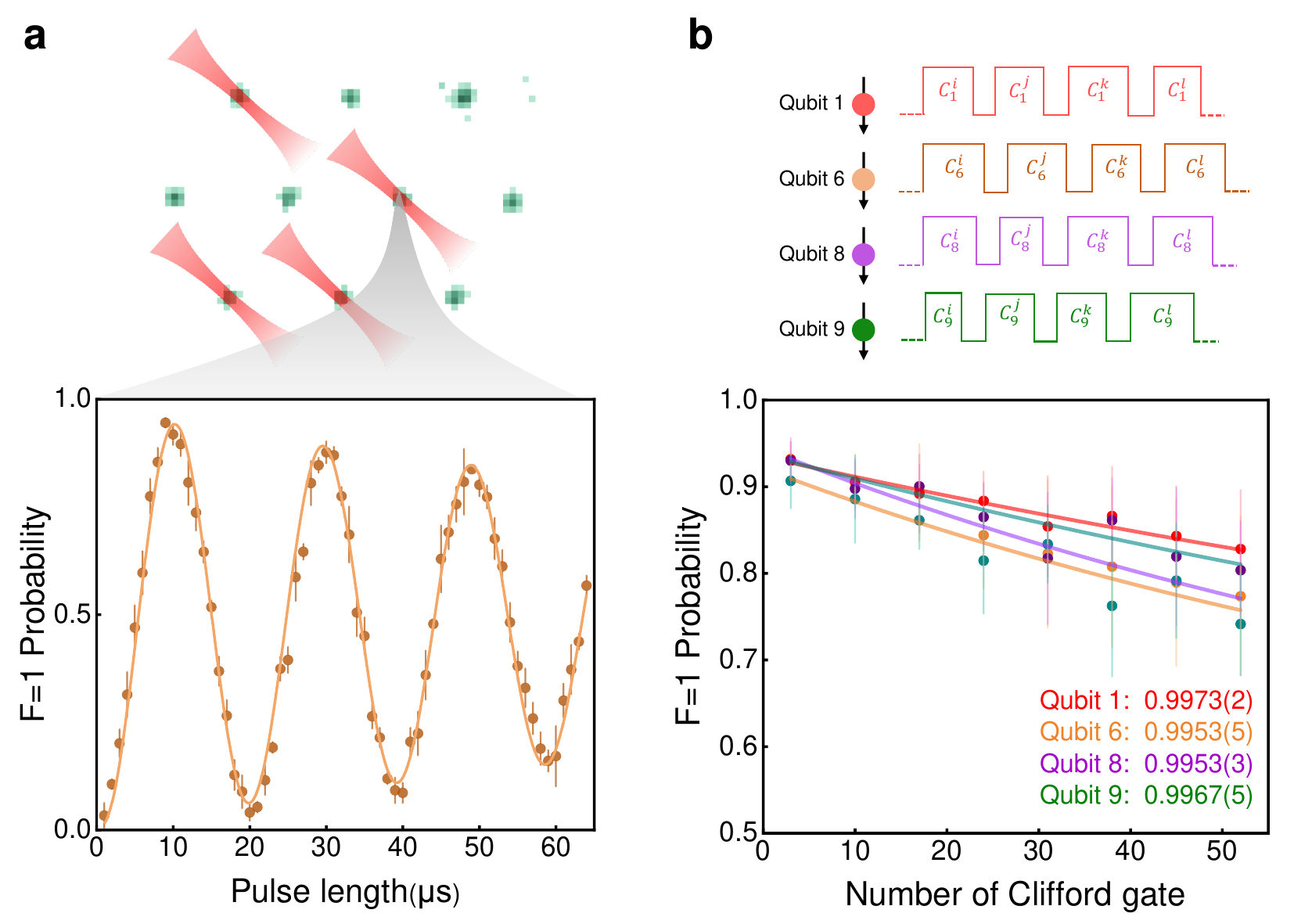}
   \caption{\textbf{Parallel RB on four arbitrary qubits.} \textbf{a}, Parallel addressing of four arbitrary single-atom qubits. Bottom: The Rabi oscillation of qubit 6, which exhibits rapid dephasing. \textbf{b}, Simultaneous RB experiments on four targeted qubits by applying an independent sequence of Clifford gates to each qubit. The fidelity of parallel-addressed single-qubit gates ranges from 0.995 to 0.997, closely approximating that of individually addressed single-qubit gates.
   \label{fig5}}
\end{figure}

Next, we perform parallel addressing of arbitrary single-atom qubits. To begin, we present parallel Rabi oscillation in arbitrary single atoms. Four addressing beams are simultaneously applied to four randomly selected single atoms, which are arranged in an irregular pattern as shown in Fig.\,\ref{fig5}\textbf{a}. By scanning the duration of the addressing laser pulses, we obtain the Rabi oscillations of the four target qubits.

The dephasing rates of the Rabi oscillations are significantly faster than anticipated, primarily due to spatial interference among the addressing lights. In our experiment, since all the addressing lights originate from a single laser, they are coherent with one another. A small amount of crosstalk among addressing beams, resulting from optical aberrations or surface scattering, will produce significant time-varying effects on the light intensity at the addressed site, as also noted in \cite{graham2022multi,radnaev2024universal}. Addressing beams corresponding to different qubits are transmitted through separate optical fibers. Random disturbances from the room environment inevitably influence these fibers, consequently introducing random phase modulations, $\phi_i(t)$, to the addressing light. This implies that the rate of intensity fluctuations, induced by interference among the addressing beams, depend on the rate of changes in room temperature or air disturbances, which is much slower than the Rabi rate but is comparable to the event-counting rate ($\sim 1 \text{ Hz}$ in our experiment). From the Rabi oscillation data, we can numerically extract that the slow modulation amplitude of the Rabi frequency ranges from $2\%$ to $5\%$. This leads to per-gate error between $3.6 \times 10^{-4}$ and $1.2 \times 10^{-3}$, which have a minimal effect on our current gate fidelity.

Then, we perform simultaneous RB experiments on four targeted qubits by applying an independent sequence of Clifford gates to each qubit. Results are displayed in Fig.\,\ref{fig5}\textbf{b}, from which the fidelity of the parallel-addressed single-qubit gates for each qubit is determined by fitting the curves to equation (\ref{eq1}). The resulting fidelities range from 0.995 to 0.997, closely approximating the fidelity of individual-addressed single-qubit gates, which confirms that the variation in the Rabi frequency is a slow modulation.

\section{Discussion}
In conclusion, we demonstrate trapping an array of 10 single atoms in a fiber-array-generated optical tweezers. Based on this platform, we further demonstrate high fidelity individual and parallel addressed single-qubit gate operations. Our ability to simultaneously implement different single-qubit gates on arbitrary qubits is extremely helpful for improving the efficiency of quantum circuit execution.

Future works will focus on improving the fidelity of individual-addressed single-qubit gates by further lowering the temperature of single atoms (using advanced cooling techniques such as Raman-sideband-cooling \cite{kaufman2012cooling} or gray-molasses-cooling \cite{brown2019gray}) and employing addressing light that has a reduced dephasing impact on hyperfine qubits. Further improvements in the fidelity of parallel-addressed single-qubit gates will, however, be limited by crosstalk. Three straightforward approaches can significantly minimize crosstalk errors: firstly, by reducing the addressing beam’s waist radius to lower the intensity at the edges of the Gaussian beam near adjacent sites; secondly, by carefully correcting optical system aberrations and reducing photon scattering on optical surfaces; and thirdly, by offsetting the carrier frequencies of addressing lights in different channels by several MHz, far above the Rabi rate, to suppress slow drifts in the Rabi rate caused by interference from addressing lights \cite{radnaev2024universal}. Furthermore, we plan to extend this approach to the addressing of two-qubit gates. For example, by using a global $420\,$nm laser and an addressing $1013\,$nm laser, we can achieve parallel, site-selective Rydberg excitations of arbitrary single ${}^{87}\mathrm{Rb}$ atoms.

In contrast to the conventional and popular methods that use SLMs \cite{barredo2016atom} or AODs \cite{endres2016atom} to create optical tweezer arrays, our approach maximizes the inherent advantages of optical tweezers in qubit addressing. For optically addressed qubits, excluding shuttling schemes, individual addressing requires a tightly focused laser beam precisely aligned with the target qubit. And single-atom qubits are obtained through the dipole potential of tightly focused trapping lasers, a feature that distinct from other optically addressed qubits such as ions and NV centers. Our method combines the free-space optical paths of the trapping and addressing beams, taking the advantage that each needs to be tightly focused. In this way, the addressing light is inherently aligned with the qubit in space. The accuracy of alignment is only determined by the aberrations of the optical system and is not affected by drifts of the optomechanical components. Additionally, although not demonstrated in this work, the 2D optical tweezers generated by our scheme can be quickly and individually turned off, which is highly beneficial for performing deep quantum circuits on a Rydberg atom array platform \cite{bluvstein2024logical,graham2022multi}.

Admittedly, this first demonstration is carried out on a small scale, the number of qubits can be increased by simply replicating the optical modules at the input end of the fiber array, which can exceed 100 based on current industrial fabrication techniques. Further scale-up can be achieved with the help of well-developed integrated photonics technology. For instance, high-bandwidth low-power-consumption lithium niobate thin-film chips can replace bulky AOM crystals to modulate the amplitude and phase of lasers \cite{wang2018integrated,celik2022high}. Chip-based WDMs with micro-ring cavities can be used to combine trapping and addressing lights \cite{shu2022microcomb,liu2024parallel}. 3D optical waveguide chips processed by femtosecond laser direct-writing can replace fiber arrays to generate a larger number of more regularly arranged optical tweezers \cite{gattass2008femtosecond,poulios2014quantum,wang2024precise,sotirova2024low}. Combining these technologies, the number of fully controllable single-atom qubits can be scaled to 1000 or beyond.

\vspace{1cm}

\textbf{Acknowledgements:} This work was supported by the National Key Research and Development Program of China under Grant No. 2021YFA1402001, the National Innovation Program for Quantum Science and Technology of China under Grant No. 2023ZD0300401, the National Natural Science Foundation of China under Grants No. 12004397, No. 12122412, No. U22A20257, No. 12121004, No. 12241410, and No. 12104464, the CAS Project for Young Scientists in Basic Research under Grant No. YSBR-055, the Key Research Program of Frontier Science of CAS under Grant No. ZDBSLY-SLH012, the Major Program (JD) of Hubei Province under Grant No. 2023BAA020.

\clearpage

\section{Methods}

\subsection{Coherence of single-atom qubits}
\begin{figure}[tbp]
  \centering
   \includegraphics[width=\linewidth]{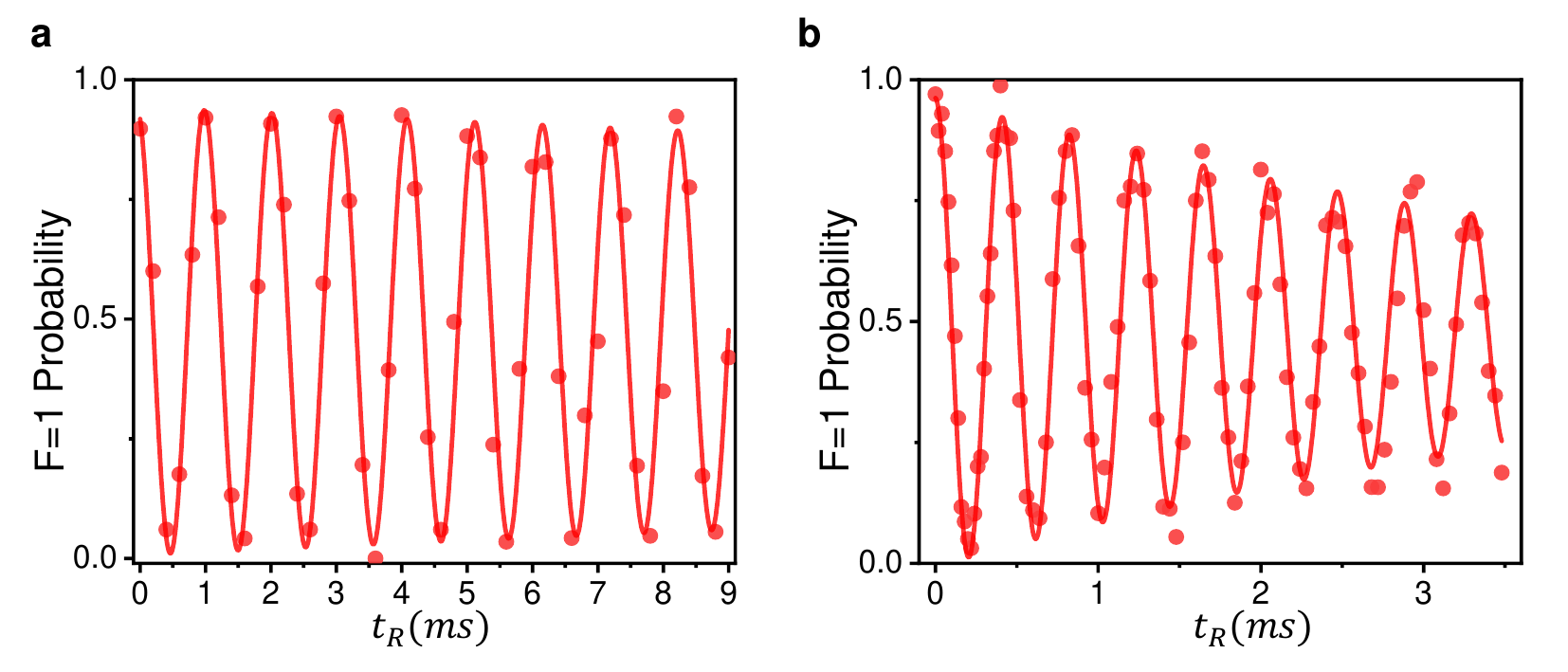}
   \caption{\textbf{Ramsey oscillations for qubit 7 in the absence (a) and presence (b) of the addressing beam.} \text Under magic-intensity trapping with $\sigma^+$ polarization, the coherence time of the qubit exceeds 50 ms. However, under the influence of a tightly focused addressing beam, the coherence time decreases to approximately 4.6 ms.
   \label{fig6}}
\end{figure}

After optically pumping the single atoms to the state $|0\rangle$, we switch the polarization of the trapping light to $\sigma^+$ using an LCVR (Thorlabs LCC1611-B), which stabilizes after 60\,ms. With this polarization configuration, single atoms are nearly trapped under magic-intensity conditions, resulting in a significantly enhanced coherence time compared to trapping with linear polarization. Under linear polarization trapping, the \(T_{2}^{*}\) coherence time of single-atom qubits is around \(3.5\,\text{ms}\). However, with magic-intensity trapping, this coherence time significantly increases to over \(50\,\text{ms}\), as indicated by the Ramsey oscillations in Fig.\,\ref{fig4}. However, a tightly focused \(795\,\)nm addressing beam (with a waist of \(2\,\mu\)m) will produce a spatially non-uniform differential AC Stark shift on the two encoded hyperfine states, thereby degrading the coherence of the qubits. Ramsey oscillations for one of the qubits, in the presence and absence of the addressing beam, are shown in  Fig.\,\ref{fig6}. Under the influence of the addressing beam, the coherence time of the qubit is reduced to approximately \(4.6\,\text{ms}\). We optimize the single-photon detuning \(\Delta\) for the two-photon Raman transitions by maximizing the product of the coherence time \(T_{2}^{*}\) and the spin-flip Rabi frequency \(\Omega_R\), yielding a value of \(200\,\text{GHz}\).

\subsection{Rabi rate Crosstalk}
\begin{figure}
   \includegraphics[width=\linewidth]{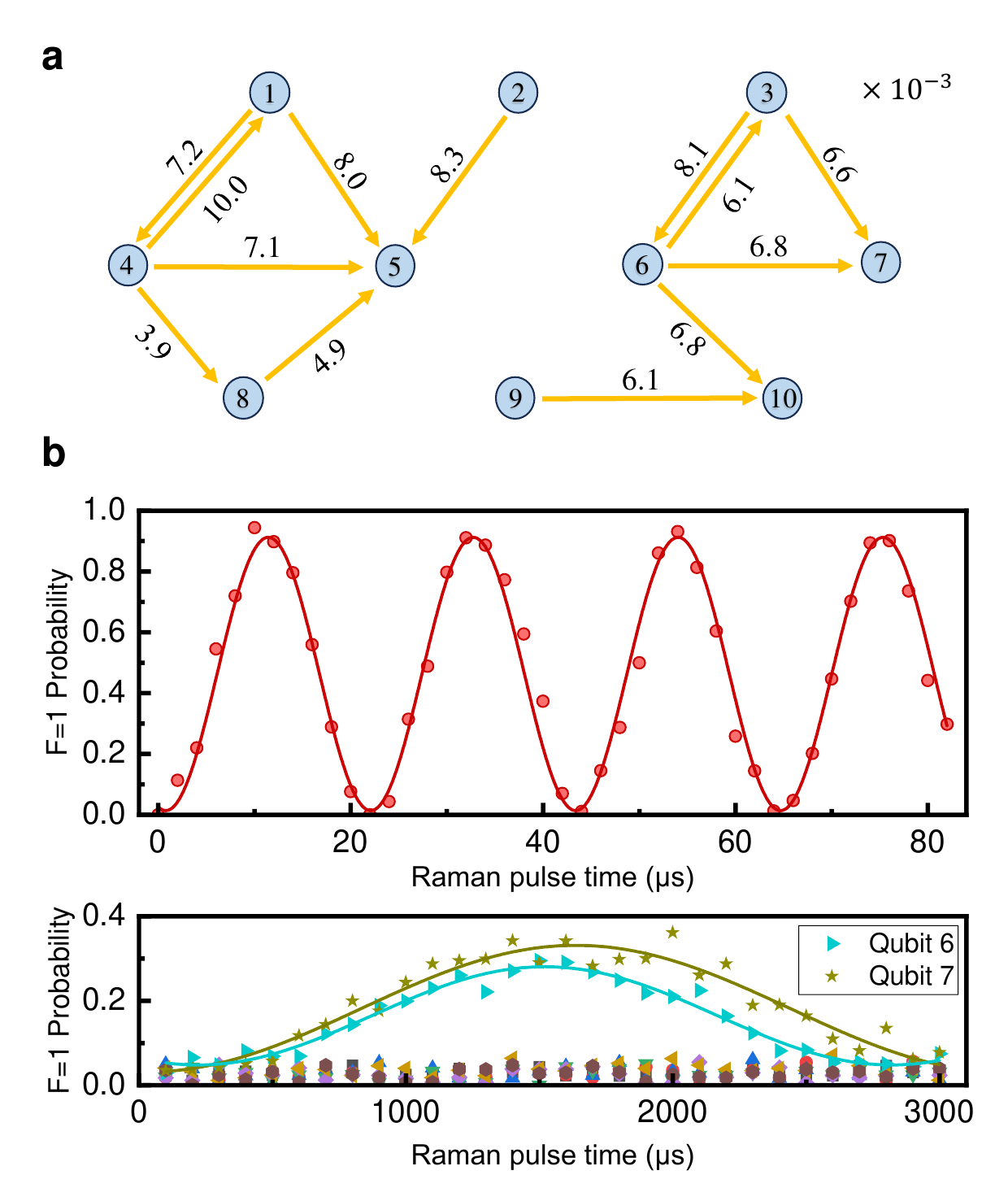}
   \caption{\textbf{Rabi rate crosstalk associated with the individual addressing of each single-atom qubit.} \textbf{a}, The overall crosstalk errors across the atom array. Each arrow originates from the addressed qubit and points towards the affected qubit, with the values representing the crosstalk rate. Qubits not linked by an arrow indicate that the crosstalk rate between them is less than 0.1\%, similar to what is shown in Fig.\,\ref{fig2}\textbf{b}, where the longest Raman pulse time used in the experiments is 30 ms. \textbf{b}, Example data of crosstalk errors when addressing qubit 3. Driven by the addressing Raman laser, qubit 3 exhibits a Rabi frequency of 47.2 kHz, whereas its neighboring qubits 6 and 7 have Rabi frequencies of 0.38 kHz and 0.31 kHz, respectively. As a result, addressing qubit 3 leads to Rabi rate crosstalk errors of 0.81\% for qubit 6 and 0.66\% for qubit 7, respectively.
   \label{fig7}}
\end{figure}

We sequentially measure the Rabi rate crosstalk for each individually addressed single atom, defined by the ratio \(\Omega_{\text{adjacent}}/\Omega_{\text{addressed}}\) (where \(\Omega_{\text{addressed}}\) and \(\Omega_{\text{adjacent}}\) respectively represent the Rabi frequencies of the addressed and adjacent single atoms). The crosstalk errors are summarized in Fig.\,\ref{fig7}. Noticeable crosstalk errors are observed between the nearest-neighbor qubits, and they display an asymmetric distribution within the array, mainly due to optical aberrations. The maximum crosstalk, measured at 1.0\%, is observed when qubit 4 is addressed, impacting its nearest-neighbor qubit 1.

\subsection{Decomposition of Clifford gate}
We decompose the 24 Clifford gate operations into two parts: rotations around axes on the equatorial plane of the Bloch sphere and rotations around the \(z\) axis. Table~\ref{table1} displays the specific decompositions of all 24 Clifford gates. The pulse phase is set to 0, \(\pi\), \(\pm \pi/2\), corresponding to the different axes \(\pm x, \pm y\) on the equatorial plane, and the pulse area corresponds to the rotation angle around these axes. The \(R_z(\theta)\) gates are virtually implemented by offsetting the phase of the addressing laser, which is equivalent to rotating the reference coordinates of subsequent pulses around the \(z\) axis by \(\theta\).

\begin{figure}
   \includegraphics[width=\linewidth]{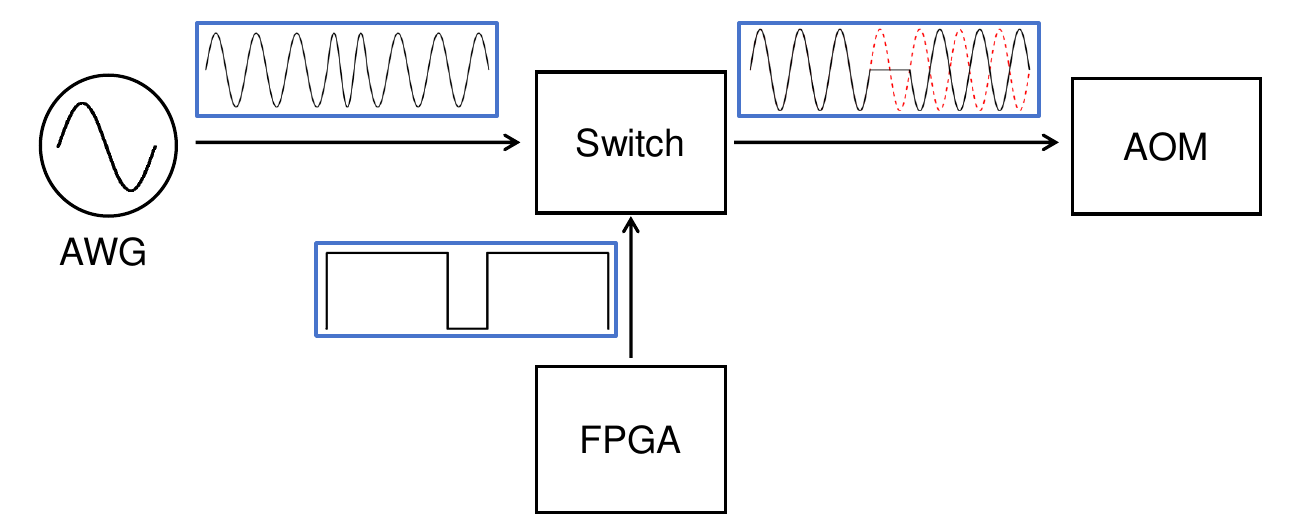}
   \caption{\textbf{The RF chain for controlling a single qubit in the RB experiment.} \text The AWG outputs a signal to drive the AOM used for controlling the addressing light, and TTL signals control the state flip of the qubit.  Phase control of the qubit is achieved by the AWG outputting an RF signal with a frequency that differs from the carrier frequency. As shown by the switch output signals, the solid black line represents the RF signal during phase switching of the qubit, while the red dashed line represents the reference signal without phase switching, indicating a shift in the initial phase of the second RF pulse relative to the first.
   \label{fig8}}
\end{figure}

In the experiment, we employ an Arbitrary Waveform Generator (AWG, Keysight M3301A) and an FPGA processor to implement the set of Clifford gates. Each addressing beam is controlled by an independent AWG channel and a TTL signal from the FPGA. All four AWG channels are employed to demonstrate parallel addressing of arbitrary single atoms. As shown in Fig.\,\ref{fig8}, the AWG outputs a carrier frequency of 110 MHz RF signal to drive the AOM used for controlling the addressing light. The FPGA processor, in turn, generates programmable TTL signals to control the pulse area of the addressing light.
An arbitrary phase offset is realized by the AWG outputting an RF signal distinct from the carrier frequency, while the addressing light keeps off during this period. For example, the AWG outputs an RF signal at \(110.5\,\text{MHz}\) lasting \(0.5\,\mu\text{s}\), which will produce a \(\pi\) phase offset in the addressing laser. To simplify the process, we keep the phase switching duration fixed and vary the phase shifts by adjusting the frequency difference.

\FloatBarrier  
\begin{table*}
\centering
\caption{\label{table1}Decompositions of Clifford gates}
\begin{ruledtabular}
\begin{tabular}{cccccccc}
Gate Index & $R_x(\theta)$ & $R_y(\theta)$ & $R_z(\theta)$ & U & Pulse Area & Pulse Phase & Phase Offset \\
\hline
0 & $I$ & $I$ & $I$ & $\begin{pmatrix} 1 & 0 \\ 0 & 1 \end{pmatrix}$ & 0 & 0 & 0 \\
1 & $I$ & $I$ & $\pi/2$ & $e^{-i\pi/4} \begin{pmatrix} 1 & 0 \\ 0 & i \end{pmatrix}$ & 0 & 0 & $\pi/2$ \\
2 & $I$ & $I$ & $\pi$ & $-i \begin{pmatrix} 1 & 0 \\ 0 & -1 \end{pmatrix}$ & 0 & 0 & $\pi$ \\
3 & $I$ & $I$ & $-\pi/2$ & $e^{i\pi/4} \begin{pmatrix} 1 & 0 \\ 0 & -i \end{pmatrix}$ & 0 & 0 & $-\pi/2$ \\
4 & $I$ & $\pi$ & $I$ & $-1 \begin{pmatrix} 0 & 1 \\ -1 & 0 \end{pmatrix}$ & $\pi$ & $\pi/2$ & 0 \\
5 & $I$ & $\pi$ & $\pi/2$ & $-e^{i\pi/4} \begin{pmatrix} 0 & 1 \\ i & 0 \end{pmatrix}$ & $\pi$ & 0 & $\pi/2$ \\
6 & $\pi$ & $I$ & $I$ & $-i \begin{pmatrix} 0 & 1 \\ 1 & 0 \end{pmatrix}$ & $\pi$ & 0 & 0 \\
7 & $\pi$ & $I$ & $\pi/2$ & $e^{-i\pi/4} \begin{pmatrix} 0 & 1 \\ -i & 0 \end{pmatrix}$ & $\pi$ & $\pi/2$ & $\pi/2$ \\
8 & $\pi$ & $\pi/2$ & $I$ & $-\frac{i}{\sqrt{2}} \begin{pmatrix} 1 & 1 \\ 1 & -1 \end{pmatrix}$ & $\pi/2$ & $-\pi/2$ & $\pi$ \\
9 & $I$ & $-\pi/2$ & $I$ & $\frac{1}{\sqrt{2}} \begin{pmatrix} 1 & 1 \\ -1 & 1 \end{pmatrix}$ & $\pi/2$ & $-\pi/2$ & 0 \\
10 & $\pi/2$ & $I$ & $\pi/2$ & $\frac{e^{-i\pi/4}}{\sqrt{2}} \begin{pmatrix} 1 & 1 \\ -i & i \end{pmatrix}$ & $\pi/2$ & $-\pi/2$ & $\pi/2$ \\
11 & $\pi/2$ & $\pi$ & $\pi/2$ & $-\frac{e^{i\pi/4}}{\sqrt{2}} \begin{pmatrix} 1 & 1 \\ i & -i \end{pmatrix}$ & $\pi/2$ & $-\pi/2$ & $-\pi/2$ \\
12 & $\pi$ & $-\pi/2$ & $I$ & $\frac{i}{\sqrt{2}} \begin{pmatrix} 1 & -1 \\ -1 & -1 \end{pmatrix}$ & $\pi/2$ & $\pi/2$ & $\pi$ \\
13 & $-\pi/2$ & $I$ & $\pi/2$ & $\frac{e^{-i\pi/4}}{\sqrt{2}} \begin{pmatrix} 1 & -1 \\ i & i \end{pmatrix}$ & $\pi/2$ & $\pi/2$ & $\pi/2$ \\
14 & $I$ & $\pi/2$ & $I$ & $\frac{1}{\sqrt{2}} \begin{pmatrix} 1 & -1 \\ 1 & 1 \end{pmatrix}$ & $\pi/2$ & $\pi/2$ & $0$ \\
15 & $-\pi/2$ & $\pi$ & $\pi/2$ & $\frac{e^{i\pi/4}}{\sqrt{2}} \begin{pmatrix} 1 & -1 \\ -i & -i \end{pmatrix}$ & $\pi/2$ & $\pi/2$ & $-\pi/2$ \\
16 & $-\pi/2$ & $-\pi/2$ & $I$ & $\frac{e^{-i\pi/4}}{\sqrt{2}} \begin{pmatrix} 1 & i \\ -1 & i \end{pmatrix}$ & $\pi/2$ & $\pi$ & $\pi/2$ \\
17 & $-\pi/2$ & $\pi/2$ & $I$ & $\frac{e^{i\pi/4}}{\sqrt{2}} \begin{pmatrix} 1 & i \\ 1 & -i \end{pmatrix}$ & $\pi/2$ & $\pi$ & $-\pi/2$ \\
18 & $-\pi/2$ & $\pi$ & $I$ & $\frac{i}{\sqrt{2}} \begin{pmatrix} 1 & i \\ -i & -1 \end{pmatrix}$ & $\pi/2$ & $\pi$ & $\pi$ \\
19 & $-\pi/2$ & $I$ & $I$ & $\frac{1}{\sqrt{2}} \begin{pmatrix} 1 & i \\ i & 1 \end{pmatrix}$ & $\pi/2$ & $\pi$ & $0$ \\
20 & $\pi/2$ & $-\pi/2$ & $I$ & $\frac{e^{i\pi/4}}{\sqrt{2}} \begin{pmatrix} 1 & -i \\ -1 & -i \end{pmatrix}$ & $\pi/2$ & $0$ & $-\pi/2$ \\
21 & $\pi/2$ & $I$ & $I$ & $\frac{1}{\sqrt{2}} \begin{pmatrix} 1 & -i \\ -i & 1 \end{pmatrix}$ & $\pi/2$ & $0$ & $0$ \\
22 & $\pi/2$ & $\pi$ & $I$ & $\frac{-i}{\sqrt{2}} \begin{pmatrix} 1 & -i \\ i & -1 \end{pmatrix}$ & $\pi/2$ & $0$ & $\pi$ \\
23 & $\pi/2$ & $\pi/2$ & $I$ & $\frac{e^{-i\pi/4}}{\sqrt{2}} \begin{pmatrix} 1 & -i \\ 1 & i \end{pmatrix}$ & $\pi/2$ & $0$ & $\pi/2$ \\
\hline
\end{tabular}
\end{ruledtabular}
\end{table*}


\begin{thebibliography}{10}
\bibliographystyle{plainnat}

\bibitem{Antoine-PRX-SLM}
Nogrette, F., Labuhn, H., Ravets, S., Barredo, D., Béguin, L., Vernier, A., Lahaye, T., Browaeys, A.
\newblock Single-atom trapping in holographic 2D arrays of microtraps with arbitrary geometries.
\newblock \emph{Physical Review X}, \textbf{4}(2):021034, 2014.

\bibitem{endres2016atom}
Endres, M., Bernien, H., Keesling, A., Levine, H., Anschuetz, E.R., Krajenbrink, A., Senko, C., Vuletic, V., Greiner, M., Lukin, M.D.
\newblock Atom-by-atom assembly of defect-free one-dimensional cold atom arrays.
\newblock \emph{Science}, \textbf{354}(6315):1024--1027, 2016.

\bibitem{barredo2016atom}
Barredo, D., De Léséleuc, S., Lienhard, V., Lahaye, T., Browaeys, A.
\newblock An atom-by-atom assembler of defect-free arbitrary two-dimensional atomic arrays.
\newblock \emph{Science}, \textbf{354}(6315):1021--1023, 2016.

\bibitem{barredo2018synthetic}
Barredo, D., Lienhard, V., De Leseleuc, S., Lahaye, T., Browaeys, A.
\newblock Synthetic three-dimensional atomic structures assembled atom by atom.
\newblock \emph{Nature}, \textbf{561}(7721):79--82, 2018.

\bibitem{schlosser2023scalable}
Schlosser, M., Tichelmann, S., Schäffner, D., de Mello, D. O., Hambach, M., Schütz, J., Birkl, G.
\newblock Scalable multilayer architecture of assembled single-atom qubit arrays in a three-dimensional Talbot tweezer lattice.
\newblock \emph{Physical Review Letters}, \textbf{130}(18):180601, 2023.

\bibitem{manetsch2024tweezer}
Manetsch, H. J., Nomura, G., Bataille, E., Leung, K. H., Lv, X., Endres, M.
\newblock A tweezer array with 6100 highly coherent atomic qubits.
\newblock \emph{arXiv preprint arXiv:2403.12021}, 2024.

\bibitem{ebadi2021quantum}
Ebadi, S., Wang, T. T., Levine, H., Keesling, A., Semeghini, G., Omran, A., Bluvstein, D., Samajdar, R., Pichler, H., Ho, W. W., \emph{et al.}
\newblock Quantum phases of matter on a 256-atom programmable quantum simulator.
\newblock \emph{Nature}, \textbf{595}(7866):227--232, 2021.

\bibitem{scholl2021quantum}
Scholl, P., Schuler, M., Williams, H. J., Eberharter, A. A., Barredo, D., Schymik, K.-N., Lienhard, V., Henry, L.-P., Lang, T. C., Lahaye, T., \emph{et al.}
\newblock Quantum simulation of 2D antiferromagnets with hundreds of Rydberg atoms.
\newblock \emph{Nature}, \textbf{595}(7866):233--238, 2021.

\bibitem{ebadi2022quantum}
Ebadi, S., Keesling, A., Cain, M., Wang, T.T., Levine, H., Bluvstein, D., Semeghini, G., Omran, A., Liu, J-G., Samajdar, R., \emph{et al.}
\newblock Quantum optimization of maximum independent set using Rydberg atom arrays.
\newblock \emph{Science}, \textbf{376}(6598):1209--1215, 2022.

\bibitem{eckner2023realizing}
Eckner, W. J., Darkwah Oppong, N., Cao, A., Young, A. W., Milner, W. R., Robinson, J. M., Ye, J., Kaufman, A. M.
\newblock Realizing spin squeezing with Rydberg interactions in an optical clock.
\newblock \emph{Nature}, \textbf{621}(7980):734--739, 2023.

\bibitem{bornet2023scalable}
Bornet, G., Emperauger, G., Chen, C., Ye, B., Block, M., Bintz, M., Boyd, J. A., Barredo, D., Comparin, T., Mezzacapo, F., \emph{et al.}
\newblock Scalable spin squeezing in a dipolar Rydberg atom array.
\newblock \emph{Nature}, \textbf{621}(7980):728--733, 2023.

\bibitem{cao2024multi}
Cao, A., Eckner, W.J., Lukin Yelin, T., Young, A.W., Jandura, S., Yan, L., Kim, K., Pupillo, G., Ye, J., Darkwah Oppong, N., \emph{et al.}
\newblock Multi-qubit gates and Schr{\"o}dinger cat states in an optical clock.
\newblock \emph{Nature}, \textbf{634}(8033):315--320, 2024.

\bibitem{finkelstein2024universal}
Finkelstein, R., Tsai, R.B.-S., Sun, X., Scholl, P., Direkci, S., Gefen, T., Choi, J., Shaw, A.L., Endres, M.
\newblock Universal quantum operations and ancilla-based read-out for tweezer clocks.
\newblock \emph{Nature}, \textbf{634}(8033):321--327, 2024.

\bibitem{browaeys2020many}
Browaeys, A., Lahaye, T.
\newblock Many-body physics with individually controlled Rydberg atoms.
\newblock \emph{Nature Physics}, \textbf{16}(2):132--142, 2020.

\bibitem{altman2021quantum}
Altman, E., Brown, K.R., Carleo, G., Carr, L.D., Demler, E., Chin, C., DeMarco, B., Economou, S.E., Eriksson, M.A., Fu, K.-M.C., \emph{et al.}
\newblock Quantum simulators: Architectures and opportunities.
\newblock \emph{PRX Quantum}, \textbf{2}(1):017003, 2021.

\bibitem{bernien2017probing}
Bernien, H., Schwartz, S., Keesling, A., Levine, H., Omran, A., Pichler, H., Choi, S., Zibrov, A.S., Endres, M., Greiner, M., \emph{et al.}
\newblock Probing many-body dynamics on a 51-atom quantum simulator.
\newblock \emph{Nature}, \textbf{551}(7682):579--584, 2017.

\bibitem{de2019observation}
De Léséleuc, S., Lienhard, V., Scholl, P., Barredo, D., Weber, S., Lang, N., Büchler, H.P., Lahaye, T., Browaeys, A.
\newblock Observation of a symmetry-protected topological phase of interacting bosons with Rydberg atoms.
\newblock \emph{Science}, \textbf{365}(6455):775--780, 2019.

\bibitem{kim2020quantum}
Kim, M., Song, Y., Kim, J., Ahn, J.
\newblock Quantum Ising Hamiltonian programming in trio, quartet, and sextet qubit systems.
\newblock \emph{PRX Quantum}, \textbf{1}(2):020323, 2020.



\bibitem{bluvstein2022quantum}
Bluvstein, D., Levine, H., Semeghini, G., Wang, T. T., Ebadi, S., Kalinowski, M., Keesling, A., Maskara, N., Pichler, H., Greiner, M., \emph{et al.}
\newblock A quantum processor based on coherent transport of entangled atom arrays.
\newblock \emph{Nature}, \textbf{604}(7906):451--456, 2022.

\bibitem{graham2022multi}
Graham, T. M., Song, Y., Scott, J., Poole, C., Phuttitarn, L., Jooya, K., Eichler, P., Jiang, X., Marra, A., Grinkemeyer, B., \emph{et al.}
\newblock Multi-qubit entanglement and algorithms on a neutral-atom quantum computer.
\newblock \emph{Nature}, \textbf{604}(7906):457--462, 2022.

\bibitem{bluvstein2024logical}
Bluvstein, D., Evered, S.J., Geim, A.A., Li, S.H., Zhou, H., Manovitz, T., Ebadi, S., Cain, M., Kalinowski, M., Hangleiter, D., \emph{et al.}
\newblock Logical quantum processor based on reconfigurable atom arrays.
\newblock \emph{Nature}, \textbf{626}(7997):58--65, 2024.

\bibitem{radnaev2024universal}
Radnaev, A.G., Chung, W.C., Cole, D.C., Mason, D., Ballance, T.G., Bedalov, M.J., Belknap, D.A., Berman, M.R., Blakely, M., Bloomfield, I.L., \emph{et al.}
\newblock A universal neutral-atom quantum computer with individual optical addressing and non-destructive readout.
\newblock \emph{arXiv preprint arXiv:2408.08288}, 2024.

\bibitem{young2024atomic}
Young, A. W., Geller, S., Eckner, W. J., Schine, N., Glancy, S., Knill, E., Kaufman, A. M.
\newblock An atomic boson sampler.
\newblock \emph{Nature}, \textbf{629}(8011):311--316, 2024.

\bibitem{yang2016coherence}
Yang, J.H., He, X.D., Guo, R.J., Xu, P., Wang, K.P., Sheng, C., Liu, M., Wang, J., Derevianko, A., Zhan, M.S.
\newblock Coherence preservation of a single neutral atom qubit transferred between magic-intensity optical traps.
\newblock \emph{Physical Review Letters}, \textbf{117}(12):123201, 2016.

\bibitem{tian2023coherence}
Tian, Z., Chang, H., Lv, X., Yang, M., Wang, Z., Yang, P., Zhang, P., Li, G., Zhang, T.C.
\newblock Coherence time of 20 s with a single cesium atom in an optical dipole trap.
\newblock \emph{arXiv preprint arXiv:2312.11196}, 2023.

\bibitem{barnes2022assembly}
Barnes, K., Battaglino, P., Bloom, B. J., Cassella, K., Coxe, R., Crisosto, N., King, J. P., Kondov, S. S., Kotru, K., Larsen, S. C., \emph{et al.}
\newblock Assembly and coherent control of a register of nuclear spin qubits.
\newblock \emph{Nature Communications}, \textbf{13}(1):2779, 2022.

\bibitem{xia2015randomized}
Xia, T., Lichtman, M., Maller, K., Carr, A.W., Piotrowicz, M.J., Isenhower, L., Saffman, M.
\newblock Randomized benchmarking of single-qubit gates in a 2D array of neutral-atom qubits.
\newblock \emph{Physical Review Letters}, \textbf{114}(10):100503, 2015.

\bibitem{wang2016single}
Wang, Y., Kumar, A., Wu, T.-Y., Weiss, D.S.
\newblock Single-qubit gates based on targeted phase shifts in a 3D neutral atom array.
\newblock \emph{Science}, \textbf{352}(6293):1562--1565, 2016.

\bibitem{sheng2018high}
Sheng, C., He, X.D., Xu, P., Guo, R.J., Wang, K.P., Xiong, Z.Y., Liu, M., Wang, J., Zhan, M.S.
\newblock High-fidelity single-qubit gates on neutral atoms in a two-dimensional magic-intensity optical dipole trap array.
\newblock \emph{Physical Review Letters}, \textbf{121}(24):240501, 2018.

\bibitem{nikolov2023randomized}
Nikolov, B., Diamond-Hitchcock, E., Bass, J., Spong, N.L.R., Pritchard, J.D.
\newblock Randomized benchmarking using nondestructive readout in a two-dimensional atom array.
\newblock \emph{Physical Review Letters}, \textbf{131}(3):030602, 2023.

\bibitem{levine2019parallel}
Levine, H., Keesling, A., Semeghini, G., Omran, A., Wang, T.T., Ebadi, S., Bernien, H., Greiner, M., Vuletić, V., Pichler, H., \emph{et al.}
\newblock Parallel implementation of high-fidelity multiqubit gates with neutral atoms.
\newblock \emph{Physical Review Letters}, \textbf{123}(17):170503, 2019.

\bibitem{fu2022high}
Fu, Z., Xu, P., Sun, Y., Liu, Y.-Y., He, X.-D., Li, X., Liu, M., Li, R.-B., Wang, J., Liu, L., and Zhan, M.S.
\newblock High-fidelity entanglement of neutral atoms via a Rydberg-mediated single-modulated-pulse controlled-phase gate.
\newblock \emph{Physical Review A}, \textbf{105}(4):042430, 2022.

\bibitem{evered2023high}
Evered, S.J., Bluvstein, D., Kalinowski, M., Ebadi, S., Manovitz, T., Zhou, H., Li, S.H., Geim, A.A., Wang, T.T., Maskara, N., \emph{et al.}
\newblock High-fidelity parallel entangling gates on a neutral-atom quantum computer.
\newblock \emph{Nature}, \textbf{622}(7982):268--272, 2023.

\bibitem{ma2023high}
Ma, S., Liu, G., Peng, P., Zhang, B., Jandura, S., Claes, J., Burgers, A.P., Pupillo, G., Puri, S., Thompson, J.D.
\newblock High-fidelity gates and mid-circuit erasure conversion in an atomic qubit.
\newblock \emph{Nature}, \textbf{622}(7982):279--284, 2023.

\bibitem{scholl2023erasure}
Scholl, P., Shaw, A.L., Tsai, R.B.-S., Finkelstein, R., Choi, J., Endres, M.
\newblock Erasure conversion in a high-fidelity Rydberg quantum simulator.
\newblock \emph{Nature}, \textbf{622}(7982):273--278, 2023.



\bibitem{beverland2022assessing}
Beverland, M.E., Murali, P., Troyer, M., Svore, K.M., Hoefler, T., Kliuchnikov, V., Low, G.H., Soeken, M., Sundaram, A., Vaschillo, A.
\newblock Assessing requirements to scale to practical quantum advantage.
\newblock \emph{arXiv preprint arXiv:2211.07629}, 2022.

\bibitem{poole2024architecture}
Poole, C., Graham, T.M., Perlin, M.A., Otten, M., Saffman, M.
\newblock Architecture for fast implementation of qLDPC codes with optimized Rydberg gates.
\newblock \emph{arXiv preprint arXiv:2404.18809}, 2024.

\bibitem{pecorari2024high}
Pecorari, L., Jandura, S., Brennen, G.K., Pupillo, G.
\newblock High-rate quantum LDPC codes for long-range-connected neutral atom registers.
\newblock \emph{arXiv preprint arXiv:2404.13010}, 2024.


\bibitem{arute2019quantum}
Arute, F., Arya, K., Babbush, R., Bacon, D., Bardin, J.C., Barends, R., Biswas, R., Boixo, S., Brandao, F.G.S.L., Buell, D.A., \emph{et al.}
\newblock Quantum supremacy using a programmable superconducting processor.
\newblock \emph{Nature}, \textbf{574}(7779):505--510, 2019.

\bibitem{wu2021strong}
Wu, Y., Bao, W.-S., Cao, S., Chen, F., Chen, M.-C., Chen, X., Chung, T.-H., Deng, H., Du, Y., Fan, D., \emph{et al.}
\newblock Strong quantum computational advantage using a superconducting quantum processor.
\newblock \emph{Physical Review Letters}, \textbf{127}(18):180501, 2021.

\bibitem{menssen2023scalable}
Menssen, A.J., Hermans, A., Christen, I., Propson, T., Li, C., Leenheer, A.J., Zimmermann, M., Dong, M., Larocque, H., Raniwala, H., \emph{et al.}
\newblock Scalable photonic integrated circuits for high-fidelity light control.
\newblock \emph{Optica}, \textbf{10}(10):1366--1372, 2023.

\bibitem{zhang2024scaled}
Zhang, B., Peng, P., Paul, A., Thompson, J.D.
\newblock Scaled local gate controller for optically addressed qubits.
\newblock \emph{Optica}, \textbf{11}(2):227--233, 2024.

\bibitem{graham2023multiscale}
Graham, T.M., Oh, E., Saffman, M.
\newblock Multiscale architecture for fast optical addressing and control of large-scale qubit arrays.
\newblock \emph{Applied Optics}, \textbf{62}(12):3242--3251, 2023.

\bibitem{ke2016tailored}
Ke, M., Zhou, F., Li, X., Wang, J., Zhan, M.S.
\newblock Tailored-waveguide based photonic chip for manipulating an array of single neutral atoms.
\newblock \emph{Optics Express}, \textbf{24}(9):9157--9167, 2016.


\bibitem{pogorelov2021compact}
Pogorelov, I., Feldker, T., Marciniak, Ch D., Postler, L., Jacob, G., Krieglsteiner, O., Podlesnic, V., Meth, M., Negnevitsky, V., Stadler, M., \emph{et al.}
\newblock Compact ion-trap quantum computing demonstrator.
\newblock \emph{PRX Quantum}, \textbf{2}(2):020343, 2021.

\bibitem{binai2023guided}
Binai-Motlagh, A., Day, M. L., Videnov, N., Greenberg, N., Senko, C., Islam, R.
\newblock A guided light system for agile individual addressing of $Ba^{+}$ qubits with $10^{-4}$ level intensity crosstalk.
\newblock \emph{Quantum Science and Technology}, \textbf{8}(4):045012, 2023.

\bibitem{sotirova2024low}
Sotirova, A.S., Sun, B., Leppard, J.D., Wang, A., Wang, M., Vazquez-Brennan, A., Nadlinger, D.P., Moser, S., Jesacher, A., He, C., \emph{et al.}
\newblock Low cross-talk optical addressing of trapped-ion qubits using a novel integrated photonic chip.
\newblock \emph{Light: Science \& Applications}, \textbf{13}(1):199, 2024.

\bibitem{flannery2024physical}
Flannery, J., Matt, R., Huber, L., Wang, K., Axline, C., Oswald, R., Home, J. P.
\newblock Physical coherent cancellation of optical addressing crosstalk in a trapped-ion experiment.
\newblock \emph{arXiv preprint arXiv:2406.06775}, 2024.

\bibitem{kobayashi2009fictitious}
Kobayashi, J., Shibata, K., Aoki, T., Kumakura, M., Takahashi, Y.
\newblock Fictitious magnetic resonance by quasielectrostatic field.
\newblock \emph{Applied Physics B}, \textbf{95}:361--365, 2009.

\bibitem{li2023low}
Li, R.-R., Chen, Y.-L., He, R., Chen, S.-Q., Qi, W.-H., Cui, J.-M., Huang, Y.-F., Li, C.-F., Guo, G.-C.
\newblock A low-crosstalk double-side addressing system using acousto-optic deflectors for atomic ion qubits.
\newblock \emph{arXiv preprint arXiv:2306.01307}, 2023.

\bibitem{kaufman2012cooling}
Kaufman, A.M., Lester, B.J., Regal, C.A.
\newblock Cooling a single atom in an optical tweezer to its quantum ground state.
\newblock \emph{Physical Review X}, \textbf{2}(4):041014, 2012.

\bibitem{brown2019gray}
Brown, M.O., Thiele, T., Kiehl, C., Hsu, T.-W., Regal, C.A.
\newblock Gray-molasses optical-tweezer loading: controlling collisions for scaling atom-array assembly.
\newblock \emph{Physical Review X}, \textbf{9}(1):011057, 2019.

\bibitem{wang2018integrated}
Wang, C., Zhang, M., Chen, X., Bertrand, M., Shams-Ansari, A., Chandrasekhar, S., Winzer, P., Lončar, M.
\newblock Integrated lithium niobate electro-optic modulators operating at CMOS-compatible voltages.
\newblock \emph{Nature}, \textbf{562}(7725):101--104, 2018.

\bibitem{celik2022high}
Celik, O.T., Sarabalis, C.J., Mayor, F.M., Stokowski, H.S., Herrmann, J.F., McKenna, T.P., Lee, N.R.A., Jiang, W., Multani, K.K.S., Safavi-Naeini, A.H.
\newblock High-bandwidth CMOS-voltage-level electro-optic modulation of 780 nm light in thin-film lithium niobate.
\newblock \emph{Optics Express}, \textbf{30}(13):23177--23186, 2022.

\bibitem{shu2022microcomb}
Shu, H., Chang, L., Tao, Y., Shen, B., Xie, W., Jin, M., Netherton, A., Tao, Z., Zhang, X., Chen, R., \emph{et al.}
\newblock Microcomb-driven silicon photonic systems.
\newblock \emph{Nature}, \textbf{605}(7910):457--463, 2022.

\bibitem{liu2024parallel}
Liu, Y., Zhang, H., Liu, J., Lu, L., Du, J., Li, Y., He, Z., Chen, J., Zhou, L., Poon, A.W.
\newblock Parallel wavelength-division-multiplexed signal transmission and dispersion compensation enabled by soliton microcombs and microrings.
\newblock \emph{Nature Communications}, \textbf{15}(1):3645, 2024.

\bibitem{gattass2008femtosecond}
Gattass, R.R., Mazur, E.
\newblock Femtosecond laser micromachining in transparent materials.
\newblock \emph{Nature Photonics}, \textbf{2}(4):219--225, 2008.

\bibitem{poulios2014quantum}
Poulios, K., Keil, R., Fry, D., Meinecke, J.D.A., Matthews, J.C.F., Politi, A., Lobino, M., Gräfe, M., Heinrich, M., Nolte, S., \emph{et al.}
\newblock Quantum walks of correlated photon pairs in two-dimensional waveguide arrays.
\newblock \emph{Physical Review Letters}, \textbf{112}(14):143604, 2014.

\bibitem{wang2024precise}
Wang, Y., Zhong, L., Lau, K.Y., Han, X., Yang, Y., Hu, J., Firstov, S., Chen, Z., Ma, Z., Tong, L., \emph{et al.}
\newblock Precise mode control of laser-written waveguides for broadband, low-dispersion 3D integrated optics.
\newblock \emph{Light: Science \& Applications}, \textbf{13}(1):130, 2024.

\end{thebibliography}
\end{document}